\documentclass[aps,showpacs,preprintnumbers,amsmath,amssymb]{revtex4}
\usepackage{graphicx,epsf}
\def\be{\begin{equation}}
\def\ee{\end{equation}}
\def\bea{\begin{eqnarray}}
\def\eea{\end{eqnarray}}

\def\S{{\cal{S}}}

\def\Dab{{\Delta_{\alpha\beta}}}

\def\m{{\mbox{\scriptsize m}}}
\def\p{{\mbox{\scriptsize nuc}}}

\def\barGamma{\Gamma_1+\Gamma_2}

\begin{document}
%
%
\title{Curvature and isocurvature perturbations in a three-fluid model
  of curvaton decay} 
\author{Sujata Gupta$^1$, Karim A.~Malik$^{2,3}$ and David Wands$^1$}
%
%
%
\affiliation{
$^1$Institute of Cosmology and Gravitation, University of Portsmouth,
Portsmouth~PO1~2EG, United Kingdom\\
$^2$GRECO, Institut d'Astrophysique de Paris, C.N.R.S.,
98bis Boulevard Arago,
75014 Paris, France\\
$^3$Physics Department, Lancaster University, 
Lancaster LA1 4YB, UK
}
\date{\today}
%
%
%
\begin{abstract}
We study the evolution of the cosmological perturbations after
inflation in curvaton models where the non-relativistic curvaton
decays into both radiation and a cold dark matter component. We
calculate the primordial curvature and correlated isocurvature
perturbations inherited by the radiation and cold dark matter
after the curvaton has decayed. We give the transfer coefficient
in terms of the initial curvaton density relative to the curvaton
decay rate. 
\end{abstract}

\pacs{98.80.Cq \hfill PU-ICG-03/43, astro-ph/0311562 v2}

\maketitle


\section{Introduction}

Observational results obtained by a variety of recent cosmic
microwave background and large scale structure experiments have
greatly improved our picture of the Universe on the largest
observable scales. These observations are consistent with a
primordial spectrum of adiabatic, Gaussian, scale-independent
perturbations~\cite{trotta,amen,lewis,wmap,gauss_test,valvi,julien,gm,sdss}.
%
An attractive explanation is that these primordial perturbations
in the early radiation-dominated era originate from quantum
fluctuations in an earlier inflationary era where small scale
vacuum fluctuations can be stretched up to arbitrarily large scales 
\cite{LLBook}.
Although it is often assumed that primordial perturbations
originate from adiabatic fluctuations of a slowly-rolling scalar
field driving the inflationary expansion (the inflaton), {\em any}
light scalar field (with effective mass less than the Hubble
scale) will acquire an almost scale-invariant spectrum during an
almost exponential expansion. Whether or not these isocurvature
fluctuations in non-inflaton fields produce a primordial
curvature perturbation depends upon whether or not the local value
of that scalar field subsequently affects the local radiation
density \cite{Mollerach,LM}.

In the curvaton scenario \cite{curvaton,MT,Enqvist} a light, weakly
coupled scalar field remains decoupled from the inflaton during
inflation and remains decoupled after the end of inflation. Once the
Hubble scale drops below the effective mass of the curvaton, the field
oscillates and eventually decays. But if the decay is late enough then
the density of the non-relativistic curvaton may have grown large
enough to contribute significantly (more than about 1\%) to the total
energy density of the universe. Then the local radiation density after
curvaton decay depends upon the local value of the curvaton at the end
of inflation and it is the initial isocurvature perturbation of the
curvaton field that determines the primordial density perturbation.
This scenario leads to new possibilities for building particle physics
models of inflation \cite{modelbuilding} and qualitatively different
constraints upon the inflationary parameters \cite{modelconstraints}.

In the conventional slow-roll inflaton scenario there is a remarkable
simplification when calculating observational constraints: one can
calculate the primordial perturbations directly in terms of the
inflaton field fluctuations during inflation, independently of the
detailed physics of the end of inflation, reheating and the subsequent
radiation-dominated era.  This is because the inflaton fluctuations
describe {\em adiabatic} perturbations about the background
homogeneous cosmology on large scales and one can define a conserved
curvature perturbation $\zeta$ directly from energy conservation
\cite{WMLL,conserved}.

By contrast, fluctuations in the curvaton field describe {\em
  non-adiabatic} perturbations about the background cosmology and the
total curvature perturbation $\zeta$ is no longer conserved on large
scales. The non-adiabatic nature of the initial perturbations leaves
open the possibility of residual isocurvature perturbations and
detectable non-Gaussianity in the curvaton scenario
\cite{curvaton,MT,LUW,cdm}. One can define individual curvature
perturbations for the radiation and curvaton fluids that remain
approximately conserved so long as the fluids can be assumed to be
non-interacting \cite{WMLL,conserved}. This has been used to derive an
approximate calculation of the resulting primordial perturbations in
the curvaton scenario in terms of the curvaton density at the decay
time in what has been called the sudden-decay approximation
\cite{curvaton,LUW}.

However an accurate calculation of the resulting primordial
perturbations requires numerical solution of the perturbation
equations. This was performed in \cite{MWU} for a two-fluid model
where non-relativistic curvaton particles decay into radiation. In
this paper we extend this analysis to a three-fluid system to
include the possibility that the curvaton also decays into
non-relativistic cold dark matter (CDM). This allows us to
calculate the residual isocurvature perturbation that may be
inherited by the matter in this scenario. See
Ref.~\cite{MT} for a numerical solution of the
perturbed curvaton equations in a field theory description.
Analytic estimates of the amplitude of the residual isocurvature
perturbation in other models for the CDM were given in
Ref.~\cite{cdm}.

In section~\ref{sec:back} we set up the coupled evolution equations
for a system of dimensionless variables describing the evolution of
the background homogeneous cosmology. We then give the coupled
evolution equations for linear density perturbations in each fluid on
uniform-curvature hypersurfaces in section \ref{sec:perteq}. In an
appendix we present equations describing the evolution of curvature
and isocurvature perturbations in manifestly gauge-invariant form and
discuss the appearance of singularities for some gauge-invariant
quantities in a system such as this that includes energy transfer
between fluids. This is seen to be due to a singular choice of
hypersurface rather than a breakdown of perturbation theory.

The primordial perturbations that result from an initial curvaton
perturbation can be characterised by a transfer coefficient that
is a unique function of a single dimensionless parameter that
specifies the initial density of the curvaton relative to the
curvaton decay rate. We give an accurate analytic approximation
(to better than 1\%) for the curvature transfer coefficient. The
same coefficient also determines non-Gaussianity of the primordial
perturbations \cite{curvaton} and, in our three-fluid model, the
amplitude of the residual isocurvature perturbation. We are able
to derive, as an exact result, a simple consistency relation
between the ratio of isocurvature to curvature perturbations and
the non-Gaussianity of the primordial perturbations, which was
previously derived only using the sudden-decay approximation.

We present our conclusions in section \ref{sec:conc}.

\section{Background equations and decay}
\label{sec:back}

In this section we will first review the governing equations of the
background quantities in a Friedman--Robertson--Walker (FRW) universe
and then find the evolution equations for the perturbations for three
interacting fluids, following Ref.~\cite{MWU}.  We begin the analysis
after the end of inflation when the vacuum energy driving
inflation has decayed into radiation. We assume that at that time the
curvaton density is much smaller than the total density, consistent
with the assumption that curvaton fluctuations are initially
isocurvature perturbations.

The radiation ($\gamma$) is a relativistic fluid with pressure
$P_\gamma=\rho_\gamma/3$. We model the other two components, the
curvaton ($\sigma$) and the cold dark matter ($m$), as non-relativistic
fluids, $P_\sigma=P_m=0$.
This is a good description of the classical field dynamics once the
Hubble rate has dropped below the effective mass of the curvaton so
that the curvaton field begins coherent oscillations.


The evolution of a flat background FRW universe is governed by the
Friedmann constraint and the continuity equation, respectively,
\bea
\label{eq:Friedmann}
H^2 &=& \frac{8\pi G}{3}\rho \,, \\
\label{eq:continuity}
\dot\rho&=&-3H\left( \rho+P\right)\,,
\eea
where the dot denotes a derivative with respect to coordinate time
$t$, $H$ is the Hubble parameter, and $\rho$ and $P$
are the total energy density and the total pressure, related to the
density and pressure of the component fluids by
\be
\sum_\alpha \rho_{\alpha} =\rho \,, \qquad
\sum_\alpha P_{\alpha} =P \,.
\ee

The continuity equation for each individual fluid in the
background can be written as \cite{KS}
\be
\label{eq:dotrhoalpha}
\dot\rho_{\alpha}
=-3H\left(\rho_{\alpha}+P_{\alpha}\right) +Q_{\alpha}\,,
\ee
where $Q_{\alpha}$ describes the energy transfer per unit time to the
$\alpha$-fluid.

We assume the curvaton can decay into both radiation and cold dark
matter, described by the two decay rates,
$\Gamma_1$ and $\Gamma_2$ respectively.
For simplicity we assume the decay rates are fixed constants.
This gives the background energy transfer as
\bea
\label{eq:defQsigma}
Q_\sigma &=&-(\barGamma)\rho_\sigma \,, \\
Q_\gamma &=&\Gamma_1\rho_\sigma \,, \\
\label{eq:defQm}
Q_{\rm{m}} &=&\Gamma_2\rho_\sigma \,.
\eea
%

The cold dark matter in this model is assumed to be a non-relativistic
product of curvaton decay, which does not interact with radiation even
at these early times. $\Gamma_2/\Gamma_1$ determines the energy
density of cold dark matter particles relative to the radiation
produced by curvaton decay. For the CDM to remain subdominant
until matter-radiation equality at $T_{\rm eq}\simeq 1$eV we require
\begin{equation}
 \frac{\Gamma_2}{\Gamma_1} \lesssim
 \left[ \frac{\Omega_{\rm{m}}}{\Omega_\gamma} \right]_{\rm decay}
 = \frac{T_{\rm eq}}{T_{\rm decay}} < 10^{-6} \, .
\end{equation}
The final inequality follows from requiring that the curvaton decays
before primordial nucleosynthesis ($T_{\rm decay}>1$MeV), after which
time any significant entropy production would spoil theoretical
predictions of the primordial abundance of light elements.

The background energy transfer must always obey the constraint
\be
\label{eq:backconstraint}
\sum_\alpha Q_{\alpha}=0 \,.
\ee

It is useful to rewrite the background equations (\ref{eq:dotrhoalpha}),
in terms of dimensionless density parameters
\be
\Omega_\sigma\equiv\frac{\rho_\sigma}{\rho}\,,\qquad
\Omega_\gamma\equiv\frac{\rho_\gamma}{\rho}\,, \qquad
\Omega_{\rm{m}}\equiv\frac{\rho_{\rm{m}}}{\rho}\,,
\ee
which gives the evolution equations
\bea
\label{eq:evolomsigma}
\Omega_\sigma' &=&
\Omega_\sigma\left(\Omega_\gamma-\frac{\barGamma}{H}
\right)\,, \\
\label{eq:evolomgamma}
\Omega_\gamma' &=&
\Omega_\sigma\frac{\Gamma_1}{H}
+\Omega_\gamma\left(\Omega_\gamma-1\right) \,, \\
\label{eq:evolomm}
\Omega_{\rm{m}}' &=&
\Omega_\sigma\frac{\Gamma_2}{H}
+\Omega_{\rm{m}}\Omega_\gamma \,,
\eea
where a dash denotes the differentiation with respect to
the number of e-foldings, $N\equiv \int Hdt$.

The Friedmann constraint (\ref{eq:Friedmann}) then yields
\be
\label{eq:omfriedmann}
\Omega_\sigma+\Omega_\gamma+\Omega_{\rm{m}}=1.
\ee

The evolution of the inverse Hubble parameter is governed by
\be
\label{eq:evolg}
\left(\frac{1}{H}\right)'
= \left(1+\frac{1}{3}\Omega_\gamma\right) \left(\frac{3}{2H}\right) \,.
\ee

Hence we have an autonomous dynamical system consisting of the four
Eqs.~(\ref{eq:evolomsigma}--\ref{eq:evolomm}), and (\ref{eq:evolg}), but
the constraint (\ref{eq:omfriedmann}) makes one of them redundant.

\subsection{Numerical solutions}

\begin{figure}
\begin{center}
\includegraphics[angle=0, width=115mm]{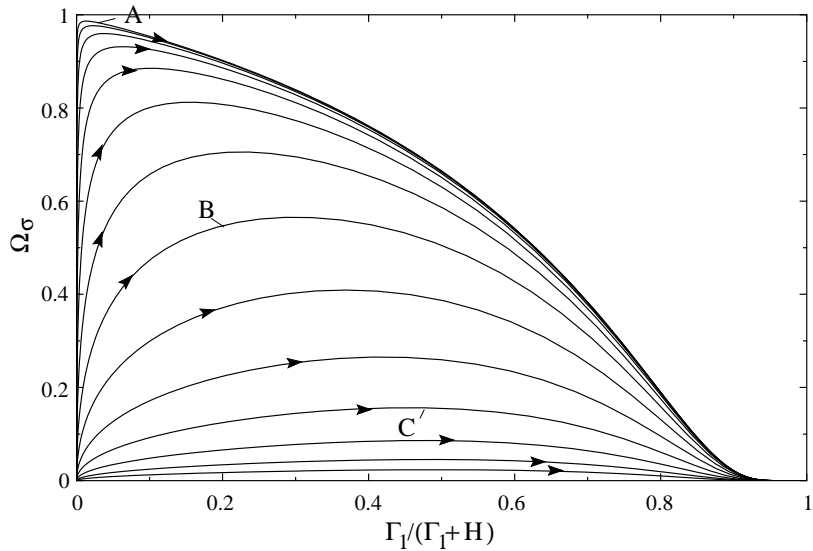}
\caption{The phase plot shows the evolution of the dimensionless
  energy density, $\Omega_\sigma$, for different values of the
  dimensionless parameter, p, defined in Eq.~(\ref{eq:defp}), varying from
  $3.15\times10^{3}$ (top line) to $3.15\times10^{-3}$ (bottom line).
  For all of the lines $\Gamma_2\ll\Gamma_1$. The evolution of the
  densities of all three fluids are plotted below in Figs.~\ref{omeg_A}
  to \ref{omeg_C} for the trajectories, marked A, B and C.}
\label{fig1}
\end{center}
\end{figure}

In Fig.\ref{fig1} we show a phase-plane which plots the evolution
of the curvaton energy density against the dimensionless decay rate,
$0<\Gamma_1/(\Gamma_1+H)<1$.
Note that for $\Gamma_2\ll\Gamma_1$ the evolution of $\Omega_\sigma$
is effectively independent of $\Gamma_2$ and the phase-plane shown is
identical to that studied in Ref.\cite{MWU}.

All trajectories in the phase-plane in Fig.\ref{fig1} start close to
the origin with $\Omega_\sigma\ll1$ and $H\gg\Gamma_1$.  {}From the
limiting form of Eqs.~(\ref{eq:evolomsigma}) and
(\ref{eq:omfriedmann}) for $\Omega_\gamma\simeq1$ we can see that
$\Omega_\sigma\propto H^{-1/2} \propto e^N$ \cite{MWU}.
Thus different trajectories can be identified by different values
of the dimensionless parameter
\be
\label{eq:defp}
p \equiv
 \left[
  \Omega_\sigma \left(\frac{H}{\;\Gamma_1}\right)^{\frac{1}{2}}
   \right]_{\mbox{\scriptsize initial}} \,.
\ee
If we consider the initial conditions for the fluid model to be set
when the curvaton begins to oscillate ($H_{\rm in}=m_\sigma$) then we
can write
\be
p = \frac{8\pi\langle \sigma^2 \rangle_{\rm in}}{3M_{\rm Pl}^2} \left(
  \frac{m_\sigma}{\Gamma_1} \right)^{1/2} \,.
\ee
The curvaton vacuum expectation value is bounded, $\langle \sigma^2
\rangle_{\rm in}<M_{\rm Pl}^2$, as we work in the regime where there
is no phase of curvaton-driven inflation~\cite{modelconstraints}. 
Nonetheless $p$ can be large if the curvaton decay rate is slow, e.g.,
$m_\sigma/\Gamma_1 \sim (M_{\rm Pl}/m_\sigma)^2$ for gravitational
strength decay.

\begin{figure}
\begin{center}
\includegraphics[angle=0, width=70mm]{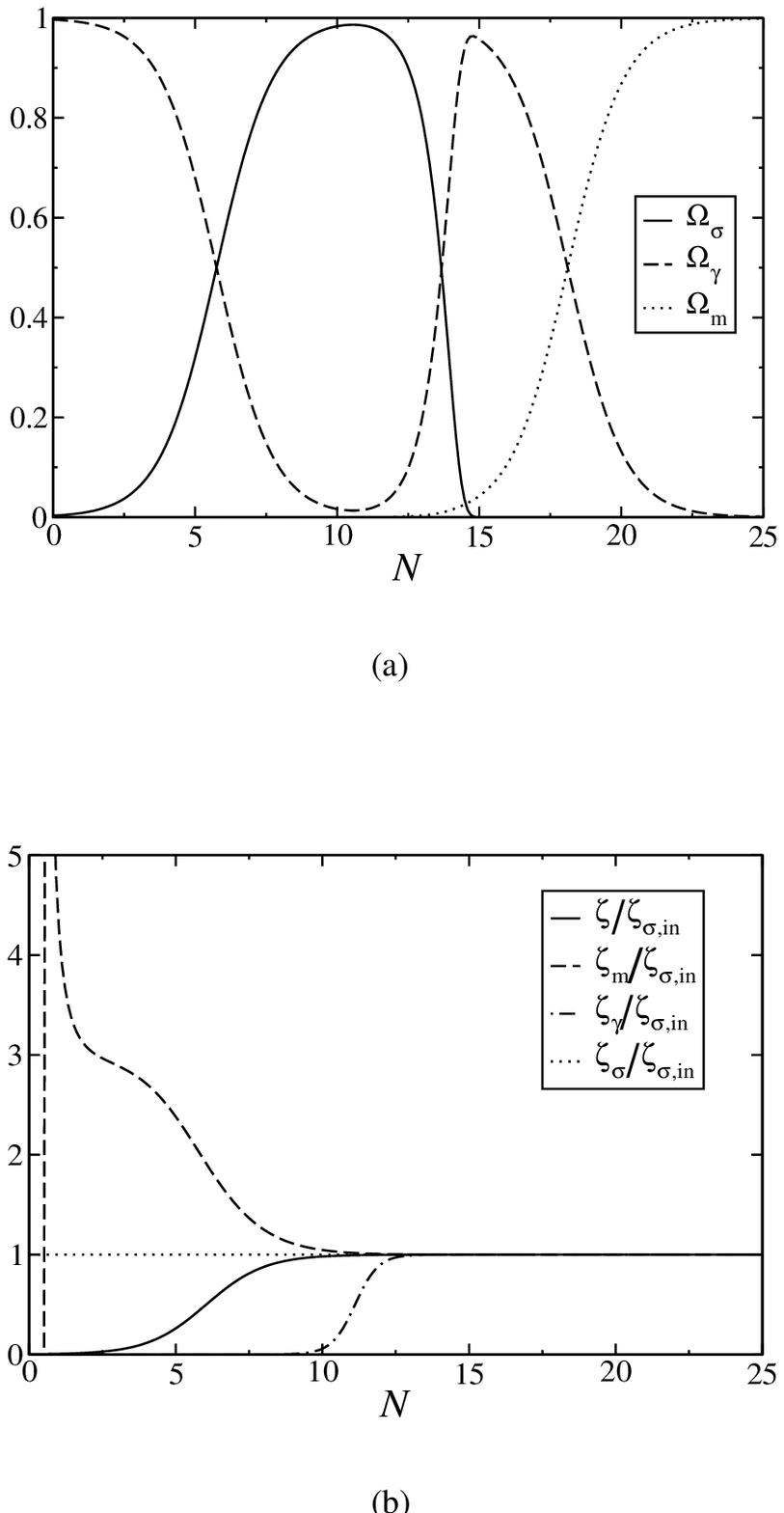}
\caption{(a) The evolution of $\Omega_\sigma$ (solid), 
$\Omega_\gamma$ (dashed) and $\Omega_{\rm{m}}$ (dotted) against
e-foldings $N$ of line A of Fig.\ref{fig1}. 
The initial value of $\Omega_\sigma$ is $10^{-2.5}$, and $\Gamma_1=10^{-10}$
 and $\Gamma_2=10^{-12}$. 
(b) The evolution of 
$\zeta_\sigma/\zeta_{\sigma {\mbox{\scriptsize ,in}}}$ (dotted), 
$\zeta_\gamma/\zeta_{\sigma {\mbox{\scriptsize ,in}}}$ (dot--dashed),
$\zeta_{\rm{m}}/\zeta_{\sigma {\mbox{\scriptsize ,in}}}$ (dashed) and 
the total curvature perturbation,
$\zeta/\zeta_{\sigma {\mbox{\scriptsize ,in}}}$ (solid), against
e-foldings $N$ of line A of Fig.\ref{fig1}.}
\label{omeg_A}
\label{zetas_A}
\end{center}
\end{figure}

\begin{figure}
\begin{center}
\includegraphics[angle=0, width=70mm]{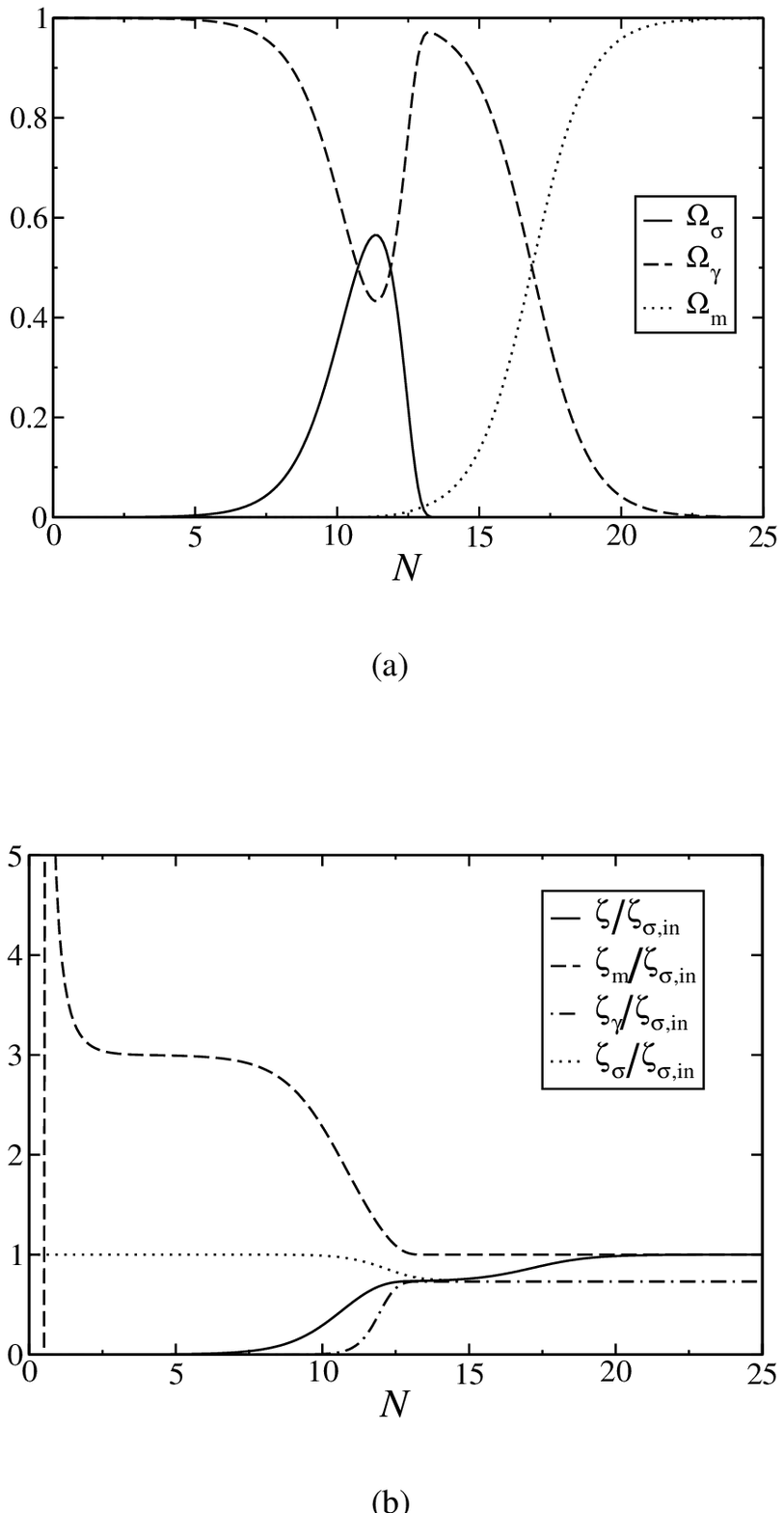}
\caption{(a) The evolution of $\Omega_\sigma$ (solid), 
$\Omega_\gamma$ (dashed) and $\Omega_{\rm{m}}$ (dotted) against
e-foldings $N$ of line B of Fig.\ref{fig1}. 
The initial value of $\Omega_\sigma$ is $10^{-4.6}$. $\Gamma_1$
 and $\Gamma_2$ are the same as in Fig.\ref{omeg_A}. 
(b) The evolution of $\zeta_\sigma/\zeta_{\sigma {\mbox{\scriptsize ,in}}}$ 
(dotted), $\zeta_\gamma/\zeta_{\sigma {\mbox{\scriptsize ,in}}}$ (dot--dashed),
$\zeta_{\rm{m}}/\zeta_{\sigma {\mbox{\scriptsize ,in}}}$ (dashed) and 
the total curvature perturbation,
$\zeta/\zeta_{\sigma {\mbox{\scriptsize ,in}}}$ (solid), against
e-foldings $N$ of line B of Fig.\ref{fig1}.}
\label{omeg_B}
\label{zetas_B}
\end{center}
\end{figure}

\begin{figure}
\begin{center}
\includegraphics[angle=0, width=70mm]{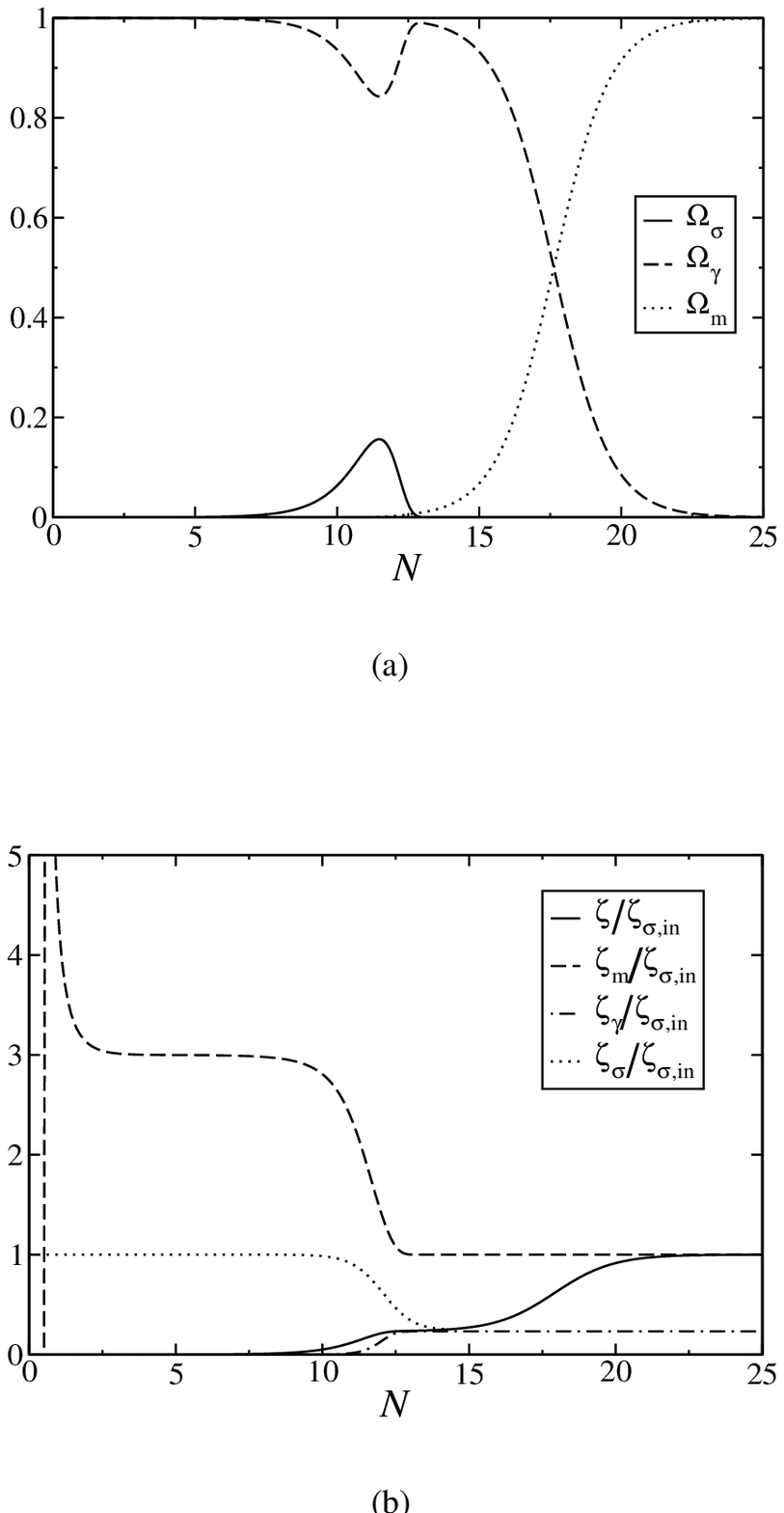}
\caption{(a) The evolution of $\Omega_\sigma$ (solid), 
$\Omega_\gamma$ (dashed) and $\Omega_{\rm{m}}$ (dotted)
against e-foldings $N$ of line C of Fig.\ref{fig1}. 
The initial value of $\Omega_\sigma$ is $10^{-5.5}$. $\Gamma_1$
 and $\Gamma_2$ are the same as in Fig.\ref{omeg_A}. 
(b) The evolution of $\zeta_\sigma/\zeta_{\sigma {\mbox{\scriptsize ,in}}}$ 
(dotted), $\zeta_\gamma/\zeta_{\sigma {\mbox{\scriptsize ,in}}}$ 
(dot--dashed), $\zeta_{\rm{m}}/\zeta_{\sigma {\mbox{\scriptsize ,in}}}$ 
(dashed) and the total curvature perturbation,
$\zeta/\zeta_{\sigma {\mbox{\scriptsize ,in}}}$ (solid), against
e-foldings $N$ of line C of Fig.\ref{fig1}.}
\label{omeg_C}
\label{zetas_C}
\end{center}
\end{figure}

We show in Figures~2 to 4 examples of the time-evolution of the three
fluid energy densities along different trajectories.
Figure~\ref{omeg_A}(a) corresponds to trajectory A of the phase--plane
Figure~\ref{fig1}. The system begins radiation dominated but as time
passes the curvaton density, with a dust equation of state, grows
relative to the radiation density. The energy density of the curvaton
then rapidly decays when $H\sim\Gamma_1$.  In this example there is a
period of curvaton domination ($\Omega_\sigma\simeq1$) before the
curvaton decays.

Figure \ref{omeg_B}(a), corresponding to trajectory B of the
phase--plane figure (\ref{fig1}), has a lower value the parameter $p$,
so the curvaton density does not climb as high before it decays away.
Figure \ref{omeg_C}(a), corresponding to trajectory C of the
phase--plane figure (\ref{fig1}), has the lowest value of $p$ of the
three plots and the ratio of initial curvaton density to the decay
rate is low enough that there is no period of curvaton domination in
this example.

\section{Perturbed equations}
\label{sec:perteq}

The evolution equation for density perturbations in the
$\alpha$ fluid is given on large scales and on uniform-curvature
hypersurfaces by \cite{MWU} 
\be
\label{eq:pertenergy2}
\dot{\delta\rho}_{\alpha}+3H(\delta\rho_{\alpha}+\delta P_{\alpha})
+Q_{\alpha}\frac{\delta\rho}{2\rho}-\delta Q_{\alpha}=0\,,
\ee
where $\delta\rho_{\alpha}$, $\delta P_{\alpha}$ and $\delta
Q_{\alpha}$ are the density perturbations, the pressure perturbation,
and the energy transfer perturbation of the $\alpha$-fluid.
$\delta\rho$ is the total density perturbation on the
uniform-curvature hypersurfaces and appears in the equation for the
perturbed energy transfer due to the gravitational time dilation
\cite{MWU}.

The perturbed energy transfer (assuming the physical decay rates are
unperturbed) is given by,
\bea
\delta Q_\sigma &=&-(\barGamma)\delta\rho_\sigma \,, \\
\delta Q_\gamma &=&\Gamma_1\delta\rho_\sigma \,, \\
\delta Q_{\rm{m}} &=&\Gamma_2\delta\rho_\sigma \,.
\label{eq:pert_trans}
\eea

We will evolve the perturbed conservation equations (\ref{eq:pertenergy2}),
which can be written in terms of the dimensionless density parameters as
%
\bea
\label{eq:del_sig}
&&\delta\rho_\sigma'+\left(3+\frac{\barGamma}{H}\right)\delta\rho_\sigma
-\frac{\barGamma}{H} \Omega_\sigma\frac{\delta\rho}{2} = 0 \,, \\
&&\delta\rho_\gamma'+4\delta\rho_\gamma
-\frac{\Gamma_1}{H}\left(\delta\rho_\sigma
-\Omega_\sigma\frac{\delta\rho}{2}\right) = 0 \,, \\
&&\delta\rho_{\rm{m}}'+3\delta\rho_{\rm{m}}
-\frac{\Gamma_2}{H}\left(\delta\rho_\sigma
-\Omega_\sigma\frac{\delta\rho}{2}\right) = 0 \,.
\label{eq:del_m}
\eea
where a prime denotes derivatives with respect to $N$.

Written in this form it is evident that the linear evolution of the
density perturbations depends only upon the dimensionless background
variables $\Omega_\alpha$ and decay rates $\Gamma_\alpha/H$. Hence the
final amplitude of the density perturbations, $\delta\rho_{\alpha,
  {\rm final}}$, relative to the initial curvaton perturbation
$\delta\rho_{\sigma, {\rm initial}}$ will depend only upon which
trajectory is followed in the background phase-space $(\Omega_\alpha,
\Gamma_\alpha/H)$. They will not depend on dimensional quantities such
as $H_{\rm initial}$. In fact we will show that the large-scale
perturbations after the curvaton has decayed can be written simply in
terms of the initial curvaton perturbation and a function of the
dimensionless parameter $p$ defined in Eq.~(\ref{eq:defp}).

\subsection{Curvature and isocurvature perturbations}

The gauge-invariant variable, $\zeta$, introduced by Bardeen
\cite{BST,Bardeen88}, is commonly used to describe the large-scale
curvature perturbation.  It is equivalent to the curvature
perturbation of uniform-density hypersurfaces and remains constant on
super--horizon scales for adiabatic perturbations.
This quantity is simply related to the total density perturbation on
uniform-curvature hypersurfaces \cite{Bardeen88,WMLL}
\be
\label{eq:zeta}
\zeta=-H\frac{\delta\rho}{\dot\rho} \,.
\ee
We can also define the curvature perturbation on uniform $\alpha$-fluid
density hypersurfaces, $\zeta_{\alpha}$, which is related to the
$\alpha$-fluid density on uniform-curvature hypersurfaces \cite{WMLL}
\be
\label{eq:zetaalpha}
\zeta_{\alpha}=-H\frac{\delta\rho_\alpha}{\dot\rho_\alpha} \,.
\ee
The total density perturbation is just the sum of the
density perturbations of the individual fluids,
$\delta\rho=\sum_\alpha\delta\rho_\alpha$,
and hence the total curvature perturbation $\zeta$ is the
weighted sum of the individual perturbations~\cite{WMLL},
\begin{equation}
\label{zetasum}
\zeta=\sum_\alpha\frac{\dot\rho_\alpha}{\dot\rho}\zeta_\alpha \,.
\end{equation}

The difference between any two individual curvature perturbations,
$\zeta_\alpha$ and $\zeta_\beta$, describes an isocurvature or entropy
perturbation~\cite{WMLL,MWU}
\begin{equation}
\label{defSab}
{\cal S}_{\alpha\beta} = 3 \left( \zeta_\alpha - \zeta_\beta \right) \,.
\end{equation}

Note that the definition of $\zeta_\alpha$ becomes singular whenever
$\dot\rho_\alpha=0$. For instance the physical density of cold dark
matter is initially vanishing, but growing due to the decay of the
curvaton ($\dot\rho_m>0$). However after the curvaton has completely
decayed, the matter density is diluted by the expansion
($\dot\rho_m<0$). At one instant we have $\dot\rho_m=0$ and $\zeta_m$
is ill-defined.
This has been previously noted in the case of an oscillating scalar
field at stationary values of the field~\cite{Brandenberger}. It is a
consequence of the uniform $\alpha$-density hypersurfaces becoming
ill-defined, and does not signal the breakdown of perturbation theory,
as long as we can find some gauge where the hypersurfaces are
well-defined and perturbations remain small (see Appendix).

\begin{figure}
\begin{center}
\includegraphics[angle=0, width=70mm]{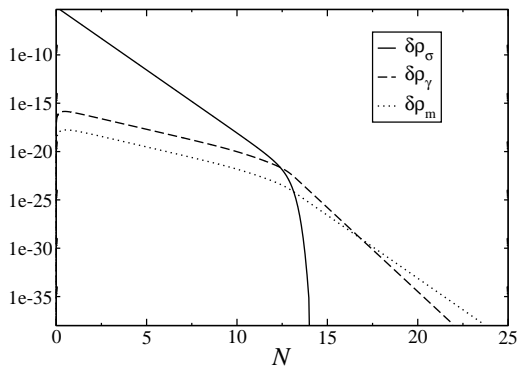}
\caption{The evolution of the density perturbations on uniform-curvature
  hypersurfaces, $\delta\rho_\sigma$ (solid), $\delta\rho_\gamma$
(dashed) and $\delta\rho_{\rm{m}}$ (dotted), against e-foldings $N$ along
the background trajectory B in Fig.\ref{fig1}.}
\label{deltas_B}
\end{center}
\end{figure}

In this case we will numerically solve for the fluid density
perturbations on uniform-curvature hypersurfaces.
Fig. \ref{deltas_B} shows an example of the evolution of the density
perturbations in equations (\ref{eq:del_sig}) to (\ref{eq:del_m})
above, showing that they do indeed remain well-defined throughout the
evolution.

{}From the density perturbations we then construct the
curvature perturbations, Eq.(\ref{eq:zeta}) and~(\ref{eq:zetaalpha}),
which gives
\be
\label{eq:drhosigma}
\zeta_\sigma=\,\frac{8\pi G}{3H}\,\frac{\delta\rho_\sigma}{\Omega_\sigma\left(3H+\barGamma\right)},
\ee
\be
\zeta_\gamma=-\frac{8\pi G}{3H}\frac{\delta\rho_\gamma}{\left(-4H\Omega_\gamma+\Gamma_1\Omega_\sigma\right)},
\ee
\be
\label{eq:drhom}
\zeta_{\rm{m}}=-\frac{8\pi G}{3H}\frac{\delta\rho_{\rm{m}}}{\left(-3H\Omega_{\rm{m}}+\Gamma_2\Omega_\sigma\right)},
\ee

At late times, after the curvaton has completely decayed the fluids are
non-interacting and have monotonically decreasing energy densities
(diluted by the expansion) so the curvature perturbations,
$\zeta_\alpha$, are well-defined, and constant on large scales.
Observations of, for example, the CMB directly constrain the amplitude
of the total curvature and isocurvature perturbations. On large
angular scales the temperature perturbations on the surface of
last-scattering are given by the Sachs-Wolfe effect~\cite{bmt,amen,lewis}
\begin{equation}
\left( \frac{\delta T}{T} \right)_{\rm lss} = \frac15 \left( - \zeta_\gamma -
  2 {\cal S}_{m\gamma} \right)_{\rm lss} \,.
\end{equation}

\subsection{Numerical results}

We have numerically evolved the coupled system of background equations
(\ref{eq:evolomsigma})--(\ref{eq:evolomm}) and~(\ref{eq:evolg}) and
first-order perturbations (\ref{eq:del_sig})--(\ref{eq:del_m}). In
each case we begin with a smooth ($\zeta_{\gamma\mbox{\scriptsize
    ,in}}=0$), dominant radiation fluid
($\Omega_{\gamma\mbox{\scriptsize ,in}}\simeq1$) and a subdominant
curvaton fluid ($\Omega_{\sigma\mbox{\scriptsize ,in}}\ll1$) with an
initial perturbation $\zeta_{\sigma\mbox{\scriptsize ,in}}\neq0$. The
cold dark matter density is initially zero
($\Omega_{\rm{m}\mbox{\scriptsize ,in}}=0$,
$\zeta_{\rm{m}\mbox{\scriptsize ,in}}=0$), and is produced solely from
curvaton decay.
The initial density perturbations on uniform-curvature hypersurfaces
are given in terms of $\zeta_{\alpha\mbox{\scriptsize ,in}}$ from
Eqs.~(\ref{eq:drhosigma})--(\ref{eq:drhom}).

In all cases we start the evolution at some high energy scale, with a
large value of the Hubble parameter, characterised by a small
initial value of the Hubble time relative to the decay time,
$\Gamma_1/H_{\mbox{\scriptsize in}}\ll1$. The subsequent evolution is
not directly affected by the precise value of
$\Gamma_1/H_{\mbox{\scriptsize in}}$, but rather by the parameter $p$
defined in Eq.~(\ref{eq:defp}) which characterises the phase-space
trajectory. In particular the final values of the
curvature and isocurvature perturbations are a function solely of
the parameter $p$.


Figures \ref{zetas_A}(b), \ref{zetas_B}(b) and \ref{zetas_C}(b) show
the evolution of the large--scale curvature perturbations for the
three fluids along three different background trajectories.

\subsubsection{Radiation}

The radiation curvature perturbation $\zeta_\gamma$ grows as the
curvaton decays, and reaches an equilibrium point in the phase-space where
$\zeta_\gamma=\zeta_\sigma$~\cite{MWU}. The final value for
$\zeta_\gamma$ relative to the initial value of
$\zeta_{\sigma\mbox{\scriptsize ,in}}$ depends upon the relative
density of the curvaton at the decay epoch,
$\Omega_{\sigma\mbox{\scriptsize ,decay}}$. This determines the
proportion of the final radiation that comes from the decay of the
perturbed curvaton.

For example, in Fig.~\ref{zetas_A}(a), corresponding to trajectory A
of Fig.~\ref{fig1}, the curvaton comes to dominate
($\Omega_\sigma\simeq1$) before the decay epoch. Thus almost all the
radiation at late times comes from the decay of the curvaton and
hence, in Fig.~\ref{zetas_A}(b), the radiation fluid at late times
inherits the initial curvaton perturbation,
$\zeta_\gamma\simeq\zeta_{\sigma\mbox{\scriptsize ,in}}$.
Compare this with Fig.~\ref{zetas_C}, corresponding to trajectory C
of the Fig.~\ref{fig1}. In this case the curvaton decays while
$\Omega_\sigma\ll1$ and the proportion of the final radiation
energy density which results from the decay of the perturbed curvaton
component is much smaller in this picture.


We define $r$ to be the ratio between the curvature perturbation of
the radiation component at the end of the calculation (when all of the
curvaton component has decayed away) to the initial curvaton
perturbation,
\be
\label{eq:defr}
\zeta_{\gamma\mbox{\scriptsize
    ,out}} = r \zeta_{\sigma\mbox{\scriptsize ,in}}\,.
\ee
We find that $r$ is solely a function of the choice of trajectory in
the background phase-space, Fig.~(\ref{fig1}). Each trajectory is
uniquely identified in terms of the initial data by the value of the
parameter $p$ defined in Eq.~(\ref{eq:defp}).

\begin{figure}
\begin{center}
\includegraphics[angle=0, width=70mm]{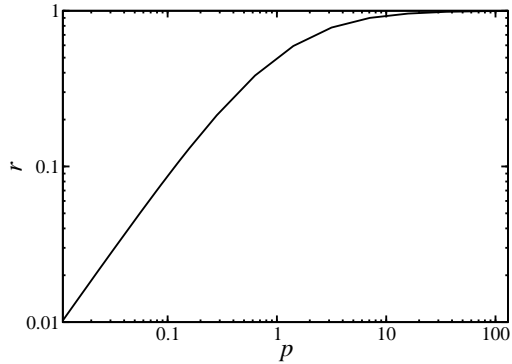}
\caption{Curvature transfer coefficient $r \equiv
\zeta_{\gamma\mbox{\scriptsize
    ,out}} / \zeta_{\sigma\mbox{\scriptsize ,in}}$,
  plotted against background parameter $p \equiv \left[
  \Omega_\sigma \left({H}/{\Gamma_1}\right)^{{1}/{2}}
   \right]_{\mbox{\scriptsize in}}$.}
\label{r_gin_p_new}
\end{center}
\end{figure}

\begin{figure}
\begin{center}
\includegraphics[angle=0, width=70mm]{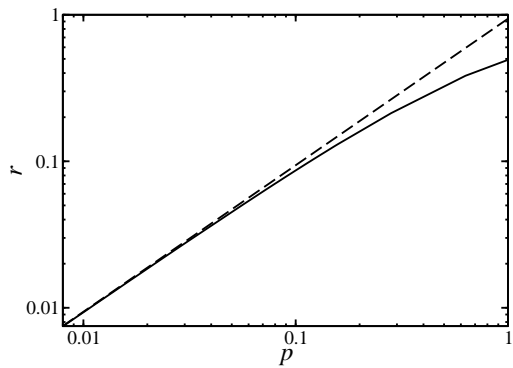}
\caption{Curvature transfer coefficient $r$, defined in
Eq.~(\ref{eq:defr}),
  plotted against parameter $p$, defined in
  Eq.~(\ref{eq:defp}), with a linear fit $r=0.924p$ for
  $p\ll1$. This linear fit is accurate to within $1\%$ for $p<0.011$.}
\label{r_gin_p_early}
\end{center}
\end{figure}

We plot $r$ against $p$ in Fig.~(\ref{r_gin_p_new}). At small values
of $p$ this approaches a linear function
\be
r \simeq 0.924 p \,,
\ee
see Fig.~(\ref{r_gin_p_early}).
This coincides with the sudden-decay approximation calculated in
Ref.~\cite{MWU} for $\Omega_{\sigma\mbox{\scriptsize ,in}}\ll1$.
The linear approximation is good to better than $1$\% up to $p=0.011$ 
($r=0.0108$) and better than $10$\% up to $p=0.11$ ($r=0.10$).

At larger values of $p$ the transfer coefficient $r(p)$ turns over
and $r\to1$ for $p\gg1$. The full function is well fit by
\be
 \label{eq:defrfit}
r_{\mbox{\scriptsize fit}}(p) = 1 - \left( 1 + \frac{0.924}{q} p
\right)^{-q} \,,
\ee
where, by fitting over the range $10^{-2}<p<10^2$, we find $q=1.24$.
This gives a fit to the numerical calculation of $r(p)$ to better
than $0.2$\% for any value of $p$.

\vspace{3mm}


\begin{figure}[ht]
\begin{center}
\includegraphics[angle=0, width=70mm]{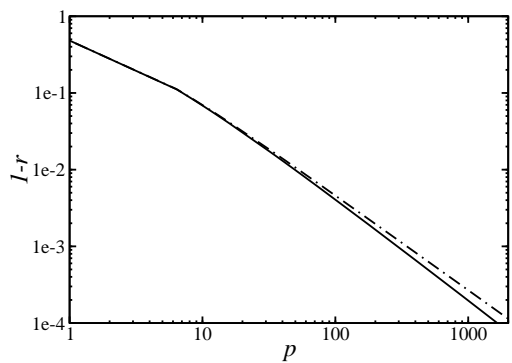}
\caption{Isocurvature transfer coefficient, $1-r$, defined in
  Eq.~(\ref{eq:defr}), plotted against parameter $p$, defined in
  Eq.~(\ref{eq:defp}). Also shown (dot-dashed line) is the fitting
  function $1-r_{\rm fit}$, defined in Eq.~(\ref{eq:defrfit}). This is
  accurate to within $1\%$ for $p<8.14$ (corresponding to
  $1-r>0.086$).}
\label{rm_gin_p}
\end{center}
\end{figure}


\subsubsection{Cold Dark Matter}
\label{sec:dust}


In our model the universe starts with an initial cold dark matter density,
$\rho_m$, equal to zero which then grows due to curvaton decay, but
eventually starts decreasing due to cosmic expansion.
As remarked earlier the curvature perturbation associated with the
cold dark matter, $\zeta_m$, becomes ill-defined when $\dot\rho_m=0$.
This occurs when $Q_m=\Gamma_2\rho_\sigma=3H\rho_m$, in
Eq.~(\ref{eq:dotrhoalpha}) for $\alpha=m$. The divergence of $\zeta_m$
is clearly seen at early times (before the curvaton decays) in Figures
\ref{zetas_A}(b), \ref{zetas_B}(b) and \ref{zetas_C}(b).

The curvature perturbation $\zeta_m$ is initially time-dependent due
to the energy transfer from the perturbed curvaton, but settles down
to a constant value at late times, once the curvaton has completely
decayed $\barGamma\gg H$.
We find that if the cold dark matter is produced solely from curvaton
decay, it inherits a curvature perturbation $\zeta_{m\mbox{\scriptsize
    ,out}}$ at late times equal to the initial curvaton perturbation,
$\zeta_{\sigma\mbox{\scriptsize ,in}}$.

This simple result can be derived by considering a composite energy density
constructed from the CDM density and a fixed fraction of the curvaton
density:
\begin{equation}
\rho_{\mbox{\scriptsize comp}} \equiv
\rho_{\rm{m}} + \frac{\Gamma_2}{\barGamma}\rho_\sigma\,.
\end{equation}
{}From Eqs.~(\ref{eq:defQsigma}) and~(\ref{eq:defQm}) we see that
the corresponding energy transfer is zero, i.e.,
\be
Q_{\mbox{\scriptsize comp}} \equiv Q_{\rm{m}} +
\frac{\Gamma_2}{\barGamma}Q_\sigma = 0\,.
\ee
Thus we obtain from Eqs.~(\ref{eq:dotrhoalpha})
and~(\ref{eq:pertenergy2}) the standard evolution
equations for the composite energy density and its
perturbation:
\begin{eqnarray}
\dot\rho_{\mbox{\scriptsize comp}}  &=& -3H\rho_{\mbox{\scriptsize
    comp}} \,,\\
\dot{\delta\rho}_{\mbox{\scriptsize comp}}
  &=& -3H\delta\rho_{\mbox{\scriptsize
    comp}} \,.
\end{eqnarray}
Note that as both the curvaton and CDM are pressureless then so is the
composite energy density, $P_{\mbox{\scriptsize comp}}=0$. Since we have
a conserved energy density with a unique equation of state the
corresponding curvature perturbation, defined by
Eq.~(\ref{eq:zetaalpha}),
\begin{equation}
\zeta_{\mbox{\scriptsize comp}}=\frac{\delta\rho_{\mbox{\scriptsize
      comp}}}{3\rho_{\mbox{\scriptsize comp}}} \,,
\end{equation}
is conserved on large scales~\cite{WMLL,conserved}.

Initially $\rho_\m=0$ and $\delta\rho_\m=0$ and hence
$\zeta_{\mbox{\scriptsize comp}}=\zeta_{\sigma\mbox{\scriptsize
    ,in}}$.
At late times
$\rho_{\rm{m}}\gg \left({\Gamma_2}/({\barGamma})\right)\rho_\sigma$,
and  $\zeta_{\mbox{\scriptsize comp}}=\zeta_{\mbox{\scriptsize m
    ,out}}$. Therefore
\be
 \label{inout}
\zeta_{\mbox{\scriptsize m ,out}}
=
\zeta_{\sigma\mbox{\scriptsize ,in}} \,,
\ee
as we see in Figs. \ref{zetas_A},  \ref{zetas_B} and \ref{zetas_C}.

\subsubsection{Curvature and isocurvature perturbations}
\label{sec:entrop}

After the curvaton has decayed, ($\Omega_\sigma\to0$),
we are left with
one overall curvature perturbation, Eq.~(\ref{zetasum}),
\be
\zeta = \frac{4\Omega_\gamma}{4\Omega_\gamma+3\Omega_{\mbox{\scriptsize m}}}
\zeta_\gamma
+ \frac{3\Omega_{\mbox{\scriptsize m}}}
{4\Omega_\gamma+3\Omega_{\mbox{\scriptsize m}}}
\zeta_{\rm{m}} \,,
\ee
and one relative isocurvature (entropy) perturbation between the
radiation and cold dark matter, Eq.~(\ref{defSab}),
\be
\mathcal{S}_{\rm{m}\gamma}=3\left({\zeta_{\rm{m}}-\zeta_\gamma}\right)
\,.
\ee

The primordial curvature perturbation is usually taken to be the
curvature perturbation at, for instance, the epoch of primordial
nucleosynthesis when the Universe is known to be radiation dominated
($\Omega_\gamma\gg\Omega_\m,\Omega_\sigma=0$), 
and hence $\zeta\simeq\zeta_\gamma$.

We have seen that the primordial curvature perturbation relative
to the initial curvaton perturbation is a function of the trajectory,
$r(p)$ in Eq.~(\ref{eq:defr}),
\be
 \left. \zeta \right|_\p \equiv \zeta_{\gamma\mbox{\scriptsize ,out}}
 = r \zeta_{\sigma\mbox{\scriptsize ,in}} \,,
\ee
whereas the final cold dark matter
curvature perturbation is identical to the initial curvaton
perturbation, Eq.~(\ref{inout}).  Hence
\be
\left. \mathcal{S} \right|_\p
\equiv \mathcal{S}_{\rm{m}\gamma\mbox{\scriptsize ,out}}
=3\left(1-r\right) \zeta_{\sigma\mbox{\scriptsize
    ,in}} \,.
\ee
If $r=1$ then we obtain an adiabatic curvature perturbation after the
curvaton has decayed and there is no late-time isocurvature
perturbation.

Note that the fitting function $r_{\rm fit}(p)$ given in
Eq.~(\ref{eq:defrfit}) gives a good approximation, for $1-r$ (better than
$1$\% for $r<0.914$ ($p<8.14$) or better than $10$\% for $r<0.993$
($p<66$)).

Therefore the primordial isocurvature and curvature perturbations are
$100\%$ correlated with their ratio fixed by $r(p)$:
\be
 \left. \frac{\mathcal{S}}{\zeta} \right|_\p =
 \frac{3\left(1-r\right)}{r} \,.
\ee
This expression was derived in Ref.~\cite{LUW} using the sudden-decay
approximation. We have shown that it is in fact an exact expression
when the CDM is produced directly from the decay of the curvaton, and
$r$ is defined by Eq.~(\ref{eq:defr}).

%

\section{Conclusions}
\label{sec:conc}

In this paper we have studied the coupled evolution of 
background
densities and linear perturbations of three interacting fluid
components: radiation, a curvaton field and cold dark matter.  We
solve numerically the evolution equations for density perturbations on
spatially flat hypersurfaces to calculate the spectrum of curvature
and isocurvature (or relative entropy) perturbations that result 
from initial perturbations in the curvaton field.

Whereas the resulting radiation curvature perturbation,
$\zeta_{\gamma,{\rm out}}$, is dependent on the initial values of background
parameters (such as curvaton density), we find the simple result that
the final CDM curvature perturbation, $\zeta_{\rm m,out}$, is always
exactly equal to the initial curvaton perturbation $\zeta_{\sigma,{\rm
    in}}$, in a model where the CDM is directly produced by curvaton
decay. We are able to demonstrate this analytically by constructing a
conserved quantity, $\zeta_{\mbox{\scriptsize comp}}$, 
that at early
times is equal to the curvaton perturbation and at late times is equal
to the CDM curvature perturbation. In effect this solves the evolution
of one degree of freedom and allows us to solve numerically only for
$\zeta_\gamma$ as a function of input parameters.

The background evolution of the fluids is described by a trajectories in
phase-space distinguished by a dimensionless parameter
$p=[\Omega(H/\Gamma_1)^{1/2}]_{\rm in}$. The ratio $r$ between
$\zeta_{\gamma,{\rm out}}$ and $\zeta_{\sigma,{\rm in}}$ 
is a function solely of this trajectory parameter $p$ and is
well-fit by a simple numerical function given in
Eq.~(\ref{eq:defrfit}). This coincides with the ``sudden-decay
approximation'' \cite{curvaton,MWU} for $p\ll1$.

The resulting radiation and CDM perturbations after curvaton decay
can then be used to predict the spectrum of CMB anisotropies in a
given model where the CDM is produced directly as a curvaton decay
product. CMB constraints are usually quoted in terms of the
primordial curvature $\zeta_\p$ and an isocurvature perturbation
$\S_\p$ around the epoch of primordial nucleosynthesis, leading to
\begin{equation}
\label{thematrix}
\left( \begin{array}{c} \zeta \\ \S \end{array} \right)_\p
= \left( \begin{array}{cc} 1 & r/3 \\ 0 & (1-r) \end{array} \right)
\left ( \begin{array}{c} \zeta \\ \S \end{array} \right)_{\rm in}
\,.
\end{equation}
In the curvaton scenario $\zeta_{\rm in}=\zeta_{\gamma,{\rm in}}=0$
and $\S_{\rm in}=3(\zeta_{\sigma,{\rm in}}-\zeta_{\gamma,{\rm
    in}})=3\zeta_{\sigma,{\rm in}}$.
Although the model we have studied describes the production of CDM
directly from the curvaton decay, one would expect a similar result to
hold for models where the net baryon or lepton number is directly
produced from curvaton decay.

The matrix equation~(\ref{thematrix}) is a particular case of the
general form of the transfer matrix for curvature and isocurvature
perturbations presented in Ref.~\cite{amen,wbmr} where the
contribution of the initial isocurvature perturbation to the
resulting primordial curvature and isocurvature perturbation is
model dependent. For instance in a curvaton scenario where the CDM
is a thermal relic which decouples from radiation sometime after
the curvaton has decayed we have instead
\begin{equation}
\label{theadmatrix}
\left( \begin{array}{c} \zeta \\ \S \end{array} \right)_\p
 = \left( \begin{array}{cc} 1 & r/3 \\ 0 & 0 \end{array} \right)
 \left ( \begin{array}{c} \zeta \\ \S \end{array} \right)_{\rm in}
\,.
\end{equation}

The matrix equations (\ref{thematrix}) and (\ref{theadmatrix})
using the numerically determined function $r(p)$ [or its
approximate form $r_{\rm fit}(p)$] enable one to relate CMB
observations 
to curvaton model parameters such as the curvaton
decay rate, $\Gamma$, and the curvaton density, $\rho_\sigma$, at some
early time, characterised by the value of the Hubble rate $H$, and
the initial spectrum of curvaton perturbations
$\zeta_\sigma=(1/3)\delta\rho_\sigma/\rho_\sigma$.

For example, in the model where the CDM is directly produced by
curvaton decay we confirm \cite{curvaton} that the relative amplitude
of primordial isocurvature to curvature perturbations is a fixed ratio
$3(1-r)/r$ which is a function solely of the background parameter $p$
defined in Eq.~(\ref{eq:defp}). This holds independently of the
``sudden-decay approximation'' previously used.

In any curvaton model the transfer coefficient
$r$ is independent of scale in the large-scale limit (by
definition) so that the scale-dependence of the primordial
perturbation spectra are inherited directly from the scale
dependence of the curvaton perturbation at the end of
inflation \cite{curvaton,wbmr} 
\begin{equation}
 \label{nsigma}
\Delta n_\sigma \equiv \frac{d\ln{\cal P}_{\delta\sigma}}{d\ln k} 
 = - 2\epsilon + 2\eta_\sigma \,.
\end{equation}
In this expression $\epsilon\equiv -\dot{H}/H^2$ is the usual
dimensionless slow-roll parameter during inflation and
$\eta_\sigma\equiv m^2_\sigma/3H^2$ is the effective mass of the
curvaton relative to the Hubble scale during inflation. Both the
curvature and isocurvature perturbation (if any) originate from
initial curvaton perturbation and are thus completely correlated,
independent of the physics of the curvaton decay.  In the absence of
any additional source for the isocurvature perturbations, the
curvature and any isocurvature perturbations must share the same
spectral tilt, given by Eq.~(\ref{nsigma}).

Note that the curvature transfer coefficient $r$ can also be
related to the non-Gaussianity of the primordial perturbations. If
$r$ is small then the initial curvaton perturbation
$\zeta_{\sigma,{\rm in}}$ must be correspondingly large in order
to produce a given amplitude of primordial curvature perturbation.
This leads to a larger contribution from second-order
perturbations in the curvaton density, which shows up in the
non-linearity parameter \cite{komsper,gauss_test}
\begin{equation}
f_{\rm NL} \approx \frac5{4r} \,,
\end{equation}
for $r\ll 1$ \cite{curvaton}. (See \cite{bmr-nongauss} for corrections
when $r\sim1$.)  There is thus a possible consistency test of this
curvaton model using two observables.  Existing constraints on the
relative amplitude of primordial isocurvature to curvature
perturbations \cite{lewis} constrain $r>0.9$ in models where the CDM
is directly produced by curvaton decay, and hence limit the allowed
non-Gaussianity.
However in curvaton models which produce purely
adiabatic perturbations ($\S_\p=0$),
determinations of the non-linear parameter $f_{\rm NL}$ may be the
only way to determine independently both the transfer parameter $r$
(and hence the model parameter $p$) and the initial amplitude of
the curvaton perturbation $\zeta_\sigma$.


\acknowledgments

The authors are grateful to David Lyth and Lorenzo Sorbo for useful
comments.  This work was supported by PPARC grants
\emph{PPA/G/S/2000/00115}, \emph{PPA/G/S/2002/00098},
and \emph{PPA/V/S/2001/00544}.  KAM was supported by a Marie Curie
Fellowship under contract number \emph{HPMF-CT-2000-00981}.  DW is
supported by the Royal Society.




\appendix
\section{Gauge-invariant adiabatic and entropy perturbations}

In a multi-component system it is often useful to re-write the
coupled perturbation equations in terms of the instantaneous
overall curvature perturbation (often called the adiabatic
perturbation) and the relative entropy perturbations between
fluids. In this appendix we follow the treatment of Ref.~\cite{MWU}
and the reader is referred to this source for more details.
Throughout this paper we work in the large scale limit, i.e.,
neglecting gradient terms.

\subsection{Evolution of curvature and entropy perturbations}

The gauge-invariant definition of the total curvature perturbation
on uniform-density hypersurfaces is~\cite{BST,Bardeen,WMLL}
\be
 \zeta \equiv -\psi - H \frac{\delta\rho}{\dot\rho} \,,
\ee
where $\psi$ and $\delta\rho$ are the gauge-dependent curvature
and density perturbations.
The evolution of the curvature perturbation is given on large
scales by~\cite{GBW,WMLL,MWU}
\be
\label{dotzetatot}
\dot\zeta = -\frac{H}{\rho+P} \delta P_{\rm nad} \,,
\ee
where the non-adiabatic pressure perturbation is $\delta P_{\rm
nad}\equiv\delta P-c_{\rm{s}}^2\delta\rho$ and the adiabatic sound
speed is $c_{\rm{s}}^2=\dot{P}/\dot\rho$. Thus the total curvature
perturbation is constant on large scales for purely adiabatic
perturbations.

In the presence of more than one fluid, the total non-adiabatic
pressure perturbation, $\delta P_{\rm nad}$, may be split into two
parts,
\be
\label{deltaPnad}
\delta P_{\rm nad}\equiv \delta P_{\rm intr}+\delta P_{\rm rel}\,.
\ee
The first part is due to the intrinsic entropy perturbation of
each fluid
\be
\label{deltaPintr}
\delta P_{\rm intr}=\sum_\alpha \delta P_{\rm{intr},\alpha} \,,
\ee
where the intrinsic non-adiabatic pressure perturbation of the
$\alpha$-fluid is given by
\be 
\label{deltaPintralpha} 
\delta P_{\rm{intr},\alpha} \equiv
\delta P_{\alpha} - c^2_{\alpha}\delta\rho_{\alpha} \,, \ee
where $c^2_{\alpha}\equiv \dot P_{\alpha}/ \dot\rho_{\alpha}$ is
the adiabatic sound speed of that fluid.
The second part of the total non-adiabatic pressure perturbation
(\ref{deltaPnad}) is due to the relative entropy perturbation
between different fluids,
\be
\label{deltaPrel}
\delta P_{\rm rel} \equiv
-\frac{1}{6H\dot\rho}
\sum_{\alpha,\beta}\dot\rho_\alpha\dot\rho_\beta
\left(c^2_\alpha-c^2_\beta\right)\S_{\alpha\beta} \,,
\ee
%
where 
$\S_{\alpha\beta}$ is the relative entropy (or isocurvature)
perturbation, Eq.~(\ref{defSab}),
\be
\label{defS}
\S_{\alpha\beta}
= -3H\left(
\frac{\delta\rho_\alpha}{\dot\rho_\alpha}
- \frac{\delta\rho_\beta}{\dot\rho_\beta} \right) \, .
\ee
Note, that Eq.~(\ref{deltaPrel}) corrects a sign error in Eq.~(2.31)
of Ref.~\cite{MWU}.

The evolution equation for the relative entropy perturbation
$\S_{\alpha\beta}$ on large scales is given by \cite{MWU}
\be
\label{S_evol}
\dot \S_{\alpha\beta}
= 3H \left(
\frac{3H\delta P_{\rm{intr},\alpha}
- \delta Q_{\rm{intr},\alpha}
- \delta Q_{\rm{rel},\alpha}}{\dot\rho_\alpha}
-\frac{3H\delta P_{\rm{intr},\beta}
- \delta Q_{\rm{intr},\beta}
- \delta Q_{\rm{rel},\beta}}{\dot\rho_\beta}\right)
 \,,
\ee
where the intrinsic non-adiabatic energy transfer perturbation
is defined as \cite{MWU}
\be \label{deltaQintralpha} \delta Q_{{\rm intr},\alpha} \equiv
\delta Q_\alpha - \frac{\dot{Q}_\alpha}{\dot\rho_\alpha}
\delta\rho_\alpha \,,
 \ee
and the relative non-adiabatic energy transfer is defined as
%
\be
\label{deltaQrelalpha}
\delta Q_{{\rm rel},\alpha} =
\frac{Q_\alpha\dot\rho}{2\rho} \left(
\frac{\delta\rho_\alpha}{\dot\rho_\alpha} - \frac{\delta\rho}{\dot\rho}
\right)
= - \frac{Q_\alpha}{6H\rho} \sum_\beta \dot\rho_\beta \S_{\alpha\beta}\,.
\ee


For fluids, such as radiation or non-relativistic matter in our
curvaton model, with definite equation of state,
$P_\alpha(\rho_\alpha)$, the intrinsic non-adiabatic pressure
perturbation in Eq.~(\ref{deltaPintr}) is zero.
Hence the total non-adiabatic pressure perturbation
(\ref{deltaPrel}) for our model can be written as
\be
\delta P_{\rm nad}
=\delta P_{\rm rel}=\frac{\dot\rho_\gamma}{9H\dot\rho} \left(
\dot\rho_{\rm m} {\cal S}_{{\rm m}\gamma} +\dot\rho_\sigma {\cal
S}_{\sigma\gamma} \right)\,.
\ee
Hence the evolution equation (\ref{dotzetatot}) for $\zeta$ is
\be
 \label{finaldotzeta}
\dot\zeta = \frac{H\dot\rho_\gamma}{3\dot\rho^2}
\left(\dot\rho_\sigma {\cal S}_{\sigma\gamma} +\dot\rho_{\rm m}
{\cal S}_{{\rm m}\gamma}\right)\,.
\ee

The intrinsic energy transfer perturbation, following from
Eq.~(\ref{eq:pert_trans}), are
 \bea
\delta Q_{\rm{intr},\sigma} &=& 0 \,,\\
\delta Q_{\rm{intr},\gamma}
 &=& -\frac{\Gamma_1}{3H}\dot\rho_\sigma{\cal S}_{\sigma\gamma} \,, \\
\delta Q_{\rm{intr,m}}
 &=& -\frac{\Gamma_2}{3H}\dot\rho_\sigma {\cal S}_{\sigma{\rm m}} \,,
 \eea
while the relative energy transfer perturbations are
\bea
\delta Q_{\rm{rel},\sigma} &=& \frac{(\barGamma) \rho_\sigma}{6H\rho}
\left(\dot\rho_\gamma {\cal S}_{\sigma\gamma}
+\dot\rho_{\rm m} {\cal S}_{\sigma{\rm m}}\right) \,, \\
\delta Q_{\rm{rel},\gamma} &=& \frac{\Gamma_1 \rho_\sigma}{6H\rho}
\left(\dot\rho_\sigma {\cal S}_{\sigma\gamma}
+\dot\rho_{\rm m} {\cal S}_{{\rm m}\gamma}\right) \,, \\
\delta Q_{\rm{rel},{\rm m}} &=& \frac{\Gamma_2 \rho_\sigma}{6H\rho}
\left(\dot\rho_\sigma {\cal S}_{\sigma{\rm m}}
-\dot\rho_{\gamma} {\cal S}_{{\rm m}\gamma}\right) \,.
\eea

Hence the evolution of the relative entropy perturbations can be
written as
\bea
 \label{Ssiggam}
\dot{\S}_{\sigma\gamma} &=& 
\frac{1}{2\rho}\left[
\Gamma_1\rho_\sigma\frac{\dot\rho_{\rm{m}}}{\dot\rho_\gamma}\S_{\rm{m}\gamma}
-(\barGamma)\rho_\sigma\frac{\dot\rho_{\rm{m}}}{\dot\rho_\sigma}\S_{\sigma\rm{m}}
-\left(2\Gamma_1\rho\frac{\dot\rho_\sigma}{\dot\rho_\gamma}
-\Gamma_2\rho_\sigma\frac{\dot\rho_\gamma}{\dot\rho_\sigma}\right)
\S_{\sigma\gamma}
\right]\,,\\
\dot{\S}_{\rm{m}\gamma} &=& \left(\Gamma_1 \frac{\dot\rho_{\rm{m}}}{\dot\rho_\gamma}
-\Gamma_2 \frac{\dot\rho_\gamma}{\dot\rho_{\rm{m}}}\right)\frac{\rho_\sigma}{2\rho}\S_{\rm{m}\gamma}
+\Gamma_2 \left(1-\frac{\rho_\sigma}{2\rho}\right)\frac{\dot\rho_\sigma}{\dot\rho_{\rm{m}}}\S_{\sigma\rm{m}}
-\Gamma_1 \left(1-\frac{\rho_\sigma}{2\rho}\right)\frac{\dot\rho_\sigma}{\dot\rho_\gamma}\S_{\sigma\gamma}\,,
\\
 \label{Ssigm}
\dot{\S}_{\sigma\rm{m}} &=&
-(\barGamma)\frac{\rho_\sigma}{2\rho}\frac{\dot\rho_\gamma}{\dot\rho_\sigma}\S_{\sigma\gamma}
-\left[\Gamma_2\left(1-\frac{\rho_\sigma}{2\rho}\right)\frac{\dot\rho_\sigma}{\dot\rho_{\rm{m}}}
+(\barGamma)\frac{\rho_\sigma}{2\rho}\frac{\dot\rho_{\rm{m}}}{\dot\rho_\sigma}\right]\S_{\sigma\rm{m}}
-\Gamma_2\frac{\rho_\sigma}{2\rho}\frac{\dot\rho_\gamma}{\dot\rho_{\rm{m}}}\S_{\rm{m}\gamma}\,.
\eea

Note that we can always identify a trivial solution to
Eqs.~(\ref{Ssiggam}--\ref{Ssigm}) with all the entropy perturbations
$\S_{\alpha\beta}=0$ which corresponds to the adiabatic mode and
leaves $\zeta=$~constant in Eq.~(\ref{finaldotzeta}), even in a system
such as this including energy transfer between different components.

\subsection{Singular and Non--Singular Hypersurfaces}


{}From Eq.~(\ref{defS}) we see that the relative entropy perturbation
$\S_{\alpha\beta}$ becomes singular whenever $\dot\rho_{\alpha}=0$.
This can occur if we have multiple fluids with energy transfer between
them whenever $Q_\alpha=3H(\rho_{\alpha}+P_{\alpha})$. This is because
the uniform $\alpha$-fluid hypersurfaces become ill-defined to
first-order whenever $\rho_\alpha$ is stationary, i.e., the
first-derivative vanishes. 

Thus a divergence in a gauge-invariant quantity such as
$\S_{\alpha\beta}$ need not signal the breakdown of perturbation
theory, but instead may indicate that a particular choice of
hypersurface was not well-defined \cite{Brandenberger}. This is indeed
the case in the curvaton model studied in this paper: we see that the density
perturbations on uniform-curvature hypersurfaces remain finite even
when $\dot\rho_{\alpha}=0$.
%


{}From the definition of ${\cal S}_{\alpha\beta}$, Eq.~(\ref{defS}) above,
we see that we can construct a non-singular quantity
\be
\Dab\equiv \dot\rho_{\alpha}\dot\rho_{\beta}
{\cal S}_{\alpha\beta}=-3H\left(\dot\rho_\beta\delta\rho_\alpha
-\dot\rho_\alpha\delta\rho_\beta\right)\,.
\ee
Since the definition of ${\cal S}_{\alpha\beta}$ is gauge invariant
then so is $\Dab$.

The evolution equation for the total curvature perturbation on large
scales, Eq.~(\ref{dotzetatot}) can then be written in terms of $\Dab$ as
\be
\label{dotzetatot2}
\dot\zeta = \frac{1}{3H(\rho +P)^2}\sum_\alpha\left\{
 H\delta P_{\rm{intr},\alpha}
-
 c^2_\alpha \sum_\beta\Dab
\right\}\,.
\ee
%

The uniform total density hypersurfaces, on which $\zeta$ is defined,
always remain well defined, since $\dot\rho\neq 0$ after inflation.

\end{document}